\documentclass[%
 aip,
 amsmath,amssymb,
 reprint,%
]{revtex4-1}

\usepackage{graphicx}
\usepackage{dcolumn}
\usepackage{bm}

\usepackage[utf8]{inputenc}
\usepackage[T1]{fontenc}
\usepackage{mathptmx}
\usepackage{etoolbox}

\makeatletter
\def\@email#1#2{%
 \endgroup
 \patchcmd{\titleblock@produce}
  {\frontmatter@RRAPformat}
  {\frontmatter@RRAPformat{\produce@RRAP{*#1\href{mailto:#2}{#2}}}\frontmatter@RRAPformat}
  {}{}
}%
\makeatother
\begin{document}

\preprint{AIP/123-QED}

\title{Mathematical modeling of the Chilean riots of 2019,\\
  an epidemiological non--local approach}
\author{Carlos Cartes}
\email{carlos.cartes@miuandes.cl}
 \affiliation{Complex Systems Group, Facultad de Ingenier\'{\i}a y Ciencias Aplicadas.\\ Universidad de los Andes, Santiago, 7620001, Chile. }

\date{\today}

\begin{abstract}

During the second half of October 2019, Chile, and especially its capital city, Santiago, suffered from widespread violence with public and private infrastructure destruction. This work aims to expand an epidemiological non--local model that successfully described the French riots of 2005 to incorporate the topology of Santiago's subway network and explain the reported distribution of rioting activity in the city. Although the model reproduced the disorders' aggregated temporal evolution, it could not deliver results resembling the observed spatial distribution of activity on Santiago. The main reason for this failure can be attributed to the fact that the model lacks a population displacement mechanism, which seems vital to explain Santiago's unrest episodes.
  
\end{abstract}

\maketitle

\begin{quotation}

A wave of generalized civil unrest started in October 2019 in Santiago, Chile. The disorders rapidly propagated to other major cities of the country, resulting in a massive destruction of public and private premises. Epidemiological models have proven helpful in describing the aggregated time evolution of a riot and reproducing the spatial distribution of events in a geographical region. This work extends one of such models, successfully used to describe the French riots of 2005, to incorporate the subway network's influence on disorder distribution, because it has proven that such network played a crucial role in the observed patterns of riot spatial density in Santiago.

\end{quotation}

\section{Introduction}

On Friday, October 18th of 2019, a generalized civil disorder started in Santiago, Chile, and, due to social media and news reports, rapidly propagated to the whole country. The incidents were triggered by an increase of $4$US cents on the transport fare in Santiago. Nevertheless, this was not the reason for the generalized riots. Instead, they were attributed to the poor quality of health, social security, and education systems, among other petitions for improving living standards. The severity of the large--scale disorder, which lasted for several days, and intermittently in some areas for months, forced the government to declare a state of emergency and impose a curfew on the country's main cities. The cost of damage to public and private property only during the first month of the activity was estimated at $4.6$billion US \cite{hribernik47countries} equivalent to $1.1\%$ of the Chilean Gross Domestic Product.

From a historical point of view, riots were called `madness of the crowds' \cite{le2017crowd} and `mass enthusiasm' \cite{lorenz2005aggression}. They were explained as a return of the collective to more primitive manners. Later works found that generalized disorder has a multifactorial origin, where poverty and a low standard of living are the principal factors explaining the severity and duration of the activity \cite{spilerman1970causes}. Qualitative approaches on the subject were introduced by the innovative work of Burbeck \cite{burbeck1978dynamics}, where an epidemiological model successfully explained the aggregated behavior of the activity registered during the racial riots in the 1960s USA and the assumption that conscious and rational decisions were involved in disorder participation \cite{granovetter1978threshold}. Epstein's formulation for social conflict \cite{epstein2002modeling} using Agent--Based Models (ABM) well represented more detailed classical patterns in public disorder, like the long calm periods interrupted by intermittent bursts of collective violence. 

Different approaches to this problem were made by using partial differential equations, capturing essential features like the contagious nature of rioting behavior \cite{petrovskii2020modelling} or the observed patterns from the French riots of 2005 \cite{berestycki2015model, bonnasse2018epidemiological}. With the same approach, Davies et al. \cite{Davies2013} successfully captured the quantitative dynamics of the London riots in 2011. This later model was based on the evidence that rational cost/benefit evaluation is involved at the moment of choice, to either join or not a riot \cite{baudains2013target}, in the same manner as regular criminal activity maximizes the potential reward by selecting different alternatives  \cite{bernasco2004residential, clare2009formal, davies2015event}. However, some evidence points out that criminal opportunism does not entirely explain rioters' target selection \cite{tiratelli2018reclaiming, drury2020social}. Besides, it was found that generalized disorder does not change baseline criminal activity \cite{mechoulan2020civil} and depends on the available means to manifest disagreement with authority \cite{boyer2020origins}.

More recently, and as a consequence of the outbreaks of violence in Chile and other Latin American countries \cite{TheConversation2019, TheGuardian2019}, Caroca et al. \cite{Caroca2020} proved that Burbeck's model adequately described the aggregated rioting activity in Santiago, Chile, during October 2019. To explain the same events, Cartes and Davies \cite{CARTESDavies} developed a numerical model that reproduces the primary spatial distribution of the riots by implementing a simple Latin American city layout. As was found by Cartes, Asahi, and Fern\'andez \cite{Cartestransport}, those disorder activities were mainly concentrated around the subway network premises and had a clear association with the residents' income. Following the same research line on the Chilean generalized unrest, we can point out an analysis based on diffusion--reaction equations, applied to country--wide economic and social variables, able to predict civil unrest due to large--scale dissatisfaction \cite{curilef2021analyzing}.

The proposition for this work is to use Bonnasse--Gahot's \cite{bonnasse2018epidemiological} model, which successfully reproduced the temporal evolution and spatial distribution of activity for the French riots of 2005, to explain the spatial distribution of the riots from Santiago during the first days of widespread violence. Because it was found that the subway network played a crucial role in promoting the formation of high activity clusters \cite{Cartestransport}. This increase of activity was already noted during the London riots of 2011 \cite{baudains2016london}, but it is also present for ordinary crime by creating awareness of new opportunities \cite{brantingham1991public, swartz2000spatial}. Therefore, public transport networks can change the way that the general population perceives distances. A similar approach was implemented by extending Davies' model for the London riots from 2011 \cite{Davies2013}, on a simplistic pattern of a Latin American city and successfully captured some basic features like the clusterization of activity in the areas with high connectivity \cite{CARTESDavies}. Bonnasse-Gahot's model was chosen over Davies' model just because of its relatively simplicity of implementation, the low number of free parameters to adjust, and the possibility of trying a new model fed with data from a major Latin American city.

This article is organized as follows, Section \ref{model} will present Bonnasse--Gahot's model, which briefly consists of space--coupled and non--local version of Burbeck's epidemiological model. Section \ref{results} will show some empirical observations from the first days of Santiago's riots, followed by some details on the numerical simulations, the extension of the model to include the effects of the subway network, and the main results. Finally, in Section \ref{discussion} a discussion and analysis of the main results are presented, with comparisons to other models and perspectives for future work.

\section{\label{model} The Model}

Burbeck's model successfully described the quantitative evolution of riots that happened in several cities in the United States during the 60's \cite{burbeck1978dynamics}. It is based on an epidemiological approach, where the number of riot events $\lambda(t)$, proportional to the number of rioters,  and the ``supply'' of riot events $\sigma(t)$, proportional to the number of available individuals \cite{Caroca2020}. Therefore the evolution equations for $\lambda$ and $\sigma$ are

\begin{eqnarray}
  \label{eq:burbeck}
  \frac{d \lambda}{dt} &=& -\omega \lambda + \beta \sigma \lambda \\
  \frac{d \sigma}{dt} &=& -\beta \sigma \lambda \nonumber
\end{eqnarray} 

\noindent where $\omega$ represents the retreat from the riot and $\beta$ the transmission rate from the supply to the disorder event. Complementing those parameters we need the initial conditions $\lambda(t_0) = \lambda_0$ and $\sigma(t_0) = \sigma_0$.

A non--local extension of this model was introduced to explain the French riots of 2005\cite{bonnasse2018epidemiological}. There Burbeck's model is generalized, to consider multiple sites therefore we have now a spatial distribution of the activity, this distribution can be performed to different spatial scales only depending on the granularity of the data.

For this work we well consider  the number of riot events at the site $k$, $\lambda_k$ and the supply at the same location $\sigma_k$. The dynamics of those quantities can be expressed as

\begin{eqnarray}
  \label{eq:nonlocal}
  \frac{d \lambda_k}{dt} &=&  -\omega_k \lambda_k + \sigma_k \Phi\left[ \Lambda_k \right]\\
  \frac{d \sigma_k}{dt} &=& -\sigma_k \Phi\left[\Lambda_k\right] \nonumber       
\end{eqnarray}

\noindent where $\Lambda_k$ is the non--local coupling which is defined by the summation

\begin{equation}
  \label{eq:coupling}
  \Lambda_k = \sum_j W_{kj}\lambda_j
\end{equation}

\noindent here the non--local kernel $W_{kj}$ has to take into account long distance interactions. For simplicity the non--local coupling function is chosen to be $\Phi\left[ \Lambda_k \right] = \beta \Lambda_k$, where the $\beta$ parameter is the same for each $k$ in the domain. Following Bonasse--Gahot indication \cite{bonnasse2018epidemiological} our kernel will have either one of the following shapes

  \begin{equation*}
  W_{kj} \sim   \left( 1 + \frac{|\vec{r}_k - \vec{r}_j|}{d_0}\right)^{-\delta}
  \end{equation*}

\noindent or

 \begin{equation*}
 W_{kj} \sim  \epsilon +  \left( 1 - \epsilon \right) e^{\frac{|\vec{r}_k - \vec{r}_j|}{d_0}}\,.
\end{equation*}

In order to make comparisons between our simulations and the model from eqns. (\ref{eq:nonlocal}), we integrated them with a fourth-order Runge-Kutta method. A wide range of parameters was explored to fit with our simulation results.

\section{\label{results} Results}

\subsection{Empirical Observations}

This section will show the main results from the data analysis belonging to the first four days of the Santiago riots in October 2019. Cartes, Asahi and Fern\'andez recently performed a complete analysis of the dataset  \cite{Cartestransport}. All the data was provided by SOSAFE, an open platform where users can report any event they found relevant. The SOSAFE platform registers the user's time, geographical coordinates, and a short description of the event. They range from relatively soft, like the temporal traffic interruption or street sign destruction, to very extreme, like the intentional burning of public and private buildings. The raw data in question is publicly available \cite{datasosafe}. The surface taken into account is delimited by Santiago's public transport system. More specifically, the geographic region is bounded by the parallels $-33.32^{\circ}$ and $-33.67^{\circ}$ latitude South and the meridians $-70.49^{\circ}$ and $-70.87^{\circ}$ longitude West.

\begin{figure}[!h]
\includegraphics{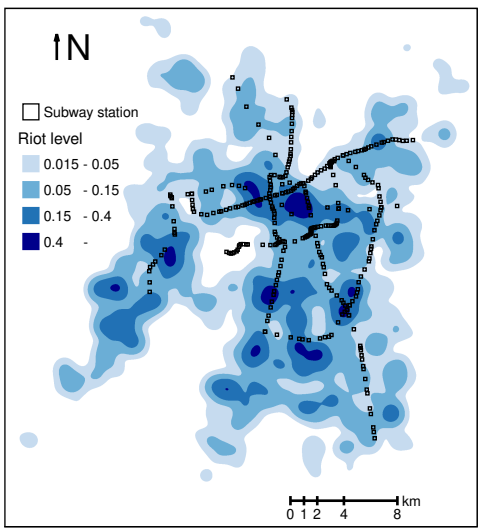}
\caption{\label{fig:riots_Santiago} Heatmap of the registered activity distribution in Santiago, from October 2019, alongside the subway network.}
\end{figure}

Data analysis reported that the temporal evolution of the disorder activity followed approximately Burbeck's model. Thus it showed an exponential growth and, after peaking, a fast decay. The activity peaked at around 22 hours, and the lowest level was at 8 hours. The spatial distribution of the riots manifested an evident clustering around the subway stations, this spatial distribution is shown in FIG. \ref{fig:riots_Santiago}, there the riot level is normalized to a scale between $0$ and $1$, being $0$ no reported activity and $1$ the maximum of all covered surface. It is important to mention that Santiago's subway network is composed by $7$ lines, $136$ stations  and $140$km, serving around $12.4\%$ of the total public transport system covering surface. Nevertheless half of the total disorders were registered at 1 km or less from the subway network and $80\%$ at 3 km or less. This correlation was found significant at the lower and higher-income neighborhoods, therefore the authors hypothesize that at least a fraction of the rioters traveled from the lower-income to the high-income regions to manifest themselves. Primarily motivated by the historical and symbolic importance of the city's most affected areas.

\subsection{Numerical Simulations}

A fourth--order Runge--Kutta method was implemented to solve the model. The space discretization was performed over a grid with dimensions $71\times 78$. Thus each element has the dimensions $500$m$\times 500$m. Thus covering the same geographic region as the aforementioned analysis \cite{Cartestransport}  and keeping the grid element size at a typical walking distance \cite{tyler2002}. Because the model needs some preliminary activity to trigger the disorder, the $20$ grid elements with the highest number of denounces were taken as the initial condition for the $\lambda$ field. This initial condition was determined with the data provided by SOSAFE. Because it is crucial to keep the relative importance between the different grid elements belonging to the initial condition, it was multiplied by a constant $\alpha < 1$.

Different kernels for the long--distance coupling and values for the numerical constants were tried. Finally, keeping the same parameters as those already tested by Bonnasse--Gahot's model was considered more practical. Because the model has already reported good agreement with the data coming from the French riots of 2005 in Paris, keeping the same parameters also makes it easier to compare the results between that model and this implementation.

The fraction of the population ``willing to join'' a riot was determined from the number of inhabitants located on each grid element, and we multiplied it by the inverse of the mean income of that grid element. Therefore it was assumed, in the same way as Davies' and Bonnasse--Gahot's models, that deprivation (which is associated to low income) is a good predictor of the population's disposition to participate in a generalized disorder. In consequence, the initial condition $\sigma(t=0)$ is defined for each grid element as the total population multiplied by its mean deprivation. It is important to mention that the deprivation's highest value is $1$ for the grid element with the lowest mean income.

The inclusion of transport networks is achieved by changing how the distance is measured between the different grid elements. In this model, distances are computed by minimizing the path between two points by using the transport network. In this case, if the distance is computed over the transport network, it is multiplied by $0.2$.
In the same way as the simplified Latin--American city pattern by Cartes and Davies\cite{CARTESDavies}.

The shortest path between two points is computed over a graph consisting of the initial and final points and all the subway stations belonging to their respective lines. Finally, the minimal path is found using the $A^*$ algorithm \cite{Hart1968}. Only the subway network was included in this work because the data showed \cite{Cartestransport} that a disproportionate large fraction of the reported disorder activity was clustered around the subway stations.

Two different kernels were tried in the simulations, exponential and power law. Results are similar in both of them. Therefore the power law kernel was used to keep the results similar to those of Bonnasse-Gahot's model. 

Finally, Table \ref{table:parameters} shows the numerical parameters used for this work. In this case, $d_0$ is chosen to be equivalent to the value $1.6\cdot 10^{-2}$km. Thus it can be comparable to the work on the French riots of 2005 \cite{bonnasse2018epidemiological}.

\begin{table}[h!]
\begin{center}
 \begin{tabular}{c | c | c } 
 
 $\omega = 0.216$ & $\beta = 4.32\cdot 10^{-4}$ & $\delta = 0.67$  \\ [0.5ex]
 \hline\hline
   $\alpha = 0.07$ & $dt = 0.05$ & $d_0 = 0.016$
 
\end{tabular}
\end{center}
\caption{Parameters used in the simulation.}
\label{table:parameters}
\end{table}

\begin{figure}[!h]
\includegraphics[width=0.5\textwidth]{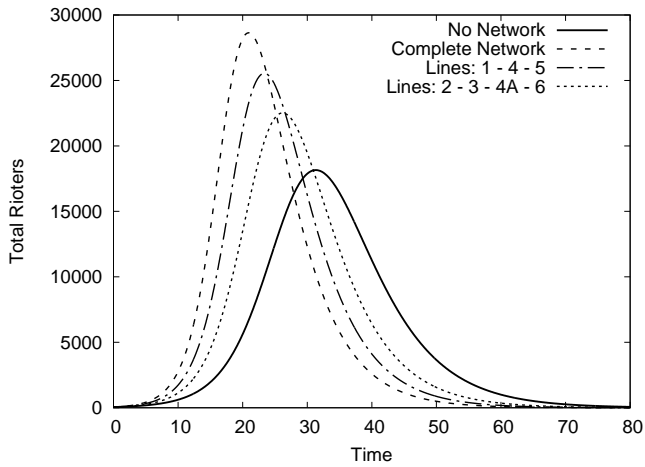}
\caption{\label{fig:total_rioters} Temporal evolution of the total number of rioters, for each transport configuration. Complete network corresponds to all seven subway lines (1, 2 , 3, 4, 4A, 5 and 6).}
\end{figure}

In FIG. \ref{fig:total_rioters} is shown the total number of rioters as function of time, $R_{\rm Total}$. for different transport network configurations. It is clear that the inclusion of a transport network increases the maximum value of $R_{\rm Total}$, turning the riot activity more intense and potentially more violent. Nevertheless, this increase in activity is also reflected by a shorter riot duration. It is a consequence of Burbeck's model due to the $\omega \lambda$ term in eqns. (\ref{eq:burbeck}). The magnitude of this increase is related to the extension of the considered transport network and the characteristics of the different connected areas of the city from FIG. \ref{fig:total_rioters}, it is clear that the largest contributor to the transport network effect is promoted by the subway lines $1$, $4$, and $5$. This is a consequence of their extension. Those three lines are composed of $80$ stations and the features of the different neighborhoods connected by those subway lines. Specifically, lines $1$, $4$ and $5$ connect the South--West parts of Santiago with the rest of the city, those areas are characterized by high population density and low--income \cite{Cartestransport}. The regions mentioned above are the larger rioter contributors. Thus, the inclusion of transport networks increases the connected areas' activation rate, which, in consequence, turns into the addition of the largest rioting population.

\begin{figure}[!h]
\includegraphics{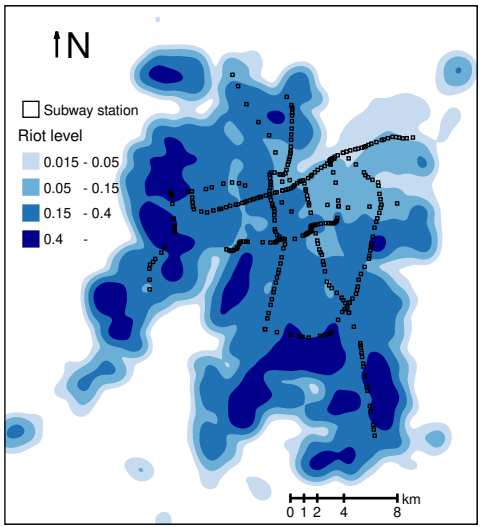}
\caption{\label{fig:beresticky_no_network} Heatmap showing the simulated activity distribution, without including subway influence, alongside the subway network.}
\end{figure}

\begin{figure}[!h]
\includegraphics{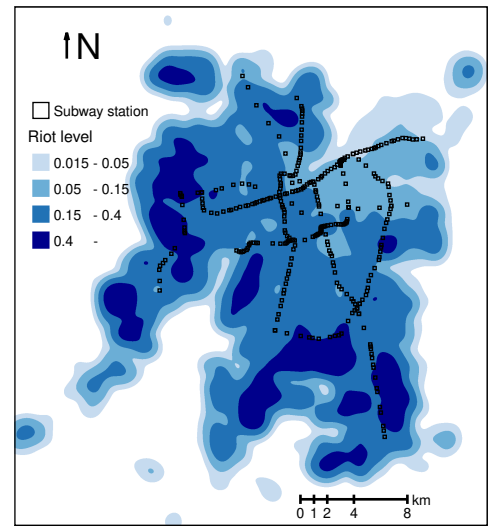}
\caption{\label{fig:beresticky_total_network} Heatmap showing the simulated activity distribution, including subway influence, alongside the subway network.}
\end{figure}

The spatial distribution of rioters, alongside the subway network, is shown on FIGS. \ref{fig:beresticky_no_network} and \ref{fig:beresticky_total_network}. In both figures, the methodology adopted was to add the rioters from all the time--steps and divide by the number of time steps. This operation will provide the average number of rioters for each grid element $k$. This result was interpolated by a gaussian kernel, with a standard deviation of $500$ meters. The intensity was normalized so that the maximum value for the rioting intensity is equal to $1$, therefore making the comparison with the actual data shown in FIG. \ref{fig:riots_Santiago} direct. In FIG. \ref{fig:beresticky_no_network} are shown the results without a transport network. There is possible to see that the regions with high rioting activity are located in areas where the population income is low \cite{Cartestransport}, and there is a relatively elevated population density. In addition, the regions with high disorder activity are more extense than those found in the real data, shown in FIG. \ref{fig:riots_Santiago}. Therefore in the absence of a transport network, most of the disorder is located in the deprived parts of the city, similar to the results found by Cartes and Davies \cite{CARTESDavies} by using Davies' model.

On the other hand, the addition of transport networks barely changed the rioters' distribution, as shown in FIG. \ref{fig:beresticky_total_network}, it can be appreciated that some of the most intense spots at the North of Santiago are slightly larger compared to FIG. \ref{fig:beresticky_no_network}, nevertheless, both heatmaps are virtually the same once they are normalized. Therefore, the inclusion of a working transport network does not affect the spatial distribution of riots in this model, even when distance perception and mutual influence are affected by it. The disorder activity stays concentrated in regions where the population density and deprivation are higher and comparatively far away from the subway network.

Finally, a different initial condition was considered, where deprivation is the same for the whole population, preserving other characteristics like density and position. Numerical simulations showed a very similar disorder distribution to those found on FIGS. \ref{fig:beresticky_no_network} and \ref{fig:beresticky_total_network}, where the most intense activity is placed where population density is higher. The only appreciable difference is some additional activity in the regions with high income, located in the city North--East. 

\section{\label{discussion} Discussion}

The epidemiological model proposed by Bonnasse--Gahot \cite{bonnasse2018epidemiological} was used to describe the riots of 2005 in France successfully. The model accurately predicted the rioting activity's disorder intensity, time evolution, and spatial distribution. Due to this success and its relatively simple implementation, the formulation above was used to replicate the first days of the Chilean riots of October 2019, specifically in the capital city Santiago. Because the model incorporates long--distance interaction mechanism, it was used to include the effects of the public transport system of Santiago on the spatial disorder distribution. In the present work, only the subway network was included because previous work \cite{Cartestransport} showed a disproportionate clustering of riot activity around Santiago's subway stations.

The direct implementation of Bonnasse--Gahot's model, without including transport network influence, showed high activity at Santiago's southwest. Coincidentally with the location of the most deprived and densely populated neighborhoods of the city. After the inclusion of transport network effects, it was found that the rioting activity increments, the peak activity and its decline occur earlier. Nevertheless, the spatial riot distribution stays practically the same, with very few changes in intensity at some areas.

It is interesting to compare these results with a similar extension implemented on Davies model \cite{Davies2013} by Cartes and Davies \cite{CARTESDavies}. In the later work, the implementation of a working transport network induced severe changes in the spatial distribution of riot intensity. But also decreasing its maximum intensity and duration, making the total disorder easier to control. Apparently, it is possible that the differences are attributed to the population displacement between different city areas, which is allowed in Davies' model but not considered in the Bonnasse--Gahot. It only permits mutual influence between the different regions. Another critical difference is that Davies' model implements targets that are attacked by the surrounding population. On the other hand, Bonnasse--Gahot's model assumes that each city region is equally attractive and prone to be attacked by its inhabitants. As a final comparison point, Davies' model showed that deprived regions presented the highest rioting activity when there was not a working transport network conducing to the same result as Bonnasse--Gahot's model.

Apparently, the possibility of displacing the population over long distances is required to describe the riot's spatial distribution observed in Santiago adequately. This behavior is not only limited to Chile's capital city but also seems similar to those observed during the "Gilets Jaunes" incidents in french cities in 2018 \cite{boyer2020origins}.

A natural extension of this work will be the implementation of Davies' model with the previous data of Santiago and try to reproduce the observed disorder distribution.

\section*{Acknowledgments}

The author wishes to acknowledge the support of  FONDECYT (CL), project number 1200357, and Universidad de los Andes (CL) through FAI initiatives.

\section*{References}

\end{document}